\documentstyle{article}
\begin{document}
\twocolumn[
\LARGE
Effects of long-range Coulomb interaction on the quantum
transport in fractional quantum Hall edges

\vspace{0.3cm}
\large
Ken-ichiro Imura and Naoto Nagaosa

\vspace{0.1cm}
\normalsize
Department of Applied Physics, University of Tokyo,
Bunkyo-ku, Tokyo 113, Japan

\vspace{0.3cm}
\small
We study the effects of long-range Coulomb
interaction (LRCI) on the quantum transport in FQH edges
with $\nu=1/(2k+1)$.
We consider two models, i.e., the
quasi-particle tunneling (QPT) model and the
electron tunneling (ET) model at the point contact.
The tunneling conductance $G(T)$
is obtained using the renormalization group treatment.
In QPT model,
it is found  that
LRCI further reduces $G(T)$ below a crossover temperature $\Lambda_w$.
In ET model, on the other hand,
there is a temeperature region
where LRCI enhances $G(T)$, and nonmonotonic temperature dependence is
predicted.

\vspace{0.1cm}

\vspace{0.3cm}
]

Quantum Hall liquid is an incompressible liquid with the gap 
in the charge excitation \cite{Zhang92}. Then the low-lying excitations are 
localized near the edge of the sample, which  
determine the low energy physics of the incompressible liquid.
These  edge modes of a fractional quantum Hall (FQH) system
are considered to be described as a chiral Tomonaga-Luttinger (TL) liquid
\cite{Wen90,Wen92}, and recent experiments seem to support this idea
showing the power law dependence of the conductance on the 
temperature and voltage \cite{Milliken96,Chang96}.
Consider a two-terminal Hall bar geometry
where the bulk FQH liquid has both upper and lower edges.
We assume that the bulk system exhibits the FQH effect with
a filling factor $\nu=1/(2k+1)$.
In this case it is expected that only one edge mode exists 
for each of the edge  when the confining potential is steep.
The edge modes for upper and lower edges have the opposite chiralities.
By applying the negative gate veltage
one can introduce the depleted region  of electrons 
sqweezing  the Hall bar.
This structure, called point contact, 
introduces interaction between the upper and lower edges, i.e., 
the backward scattering between the edges due to the quasiparticle 
tunnneling (QPT) through the bulk FQH liquid. 
This can be described by a TL model with a scattering potential 
at $x=0$ (QPT model) \cite{Kane92,Furusaki93}.
This  model predicts a low temperature 
tunneling conductance as  $ G(T) \sim T^{2/\nu-2}$
\cite{Kane92,Furusaki93,Wen91,Moon93,Fendley95},
which is consistent with the recent experiment
\cite{Milliken96}.
Another model is the electron tunneling (ET) model, where the 
depleted region is considered to be a vacuum, and the 
electron can tunnel through this region between the 
left and right FQH liquids.
This model also predicts $ G(T) \sim T^{2/\nu-2}$.
Recently, however, 
Moon and Girvin (MG) pointed out a
descrepancy between the above theory and the experiments
at very low temperatures \cite{Moon96}. 
They propose that this discrepancy is resolved
by incorpolating the effects of long-range Coulomb interaction (LRCI)
in the ET model \cite{Moon96}.

In this paper, we study the effects of LRCI 
on the quantum transport \cite{Nagaosa94,Oreg95,Sassetti97}
in the above two models of FQH 
edges.
We obtain the tunneling conductance $G(T)$ through the potential barrier
using the renormalization group treatment,
and show that QPT and ET models give qualitatively
different behaviors for the low temperature conductance, which depend
on the length scale of the system. 
In the following we employ the unit where $\hbar=k_B=1$.

\noindent
{\it Quasiparticle tunneling model} -
The model describes a two-terminal Hall bar geometry where
a two-\\
dimensional electron system between the left and right terminals
has upper and lower edges with a scattering potential at $x=0$.
\begin{equation}
S=S_0+S_a+S_w+u\int d\tau\cos\phi_+(\tau,x=0) 
\end{equation}
The first term $S_0$ describes the usual chiral TL 
liquid \cite{Wen90,Wen92,Imura97},
\[
{\cal L}_0 = {v_R\over 8\pi\nu}
\left\{
\left({\partial\phi_+ \over\partial x}\right)^2+
\left({\partial\phi_- \over\partial x}\right)^2
\right\}
+ {i\over 4\pi\nu}
{\partial\phi_+ \over\partial \tau}
{\partial\phi_- \over\partial x}
\]
where the short-range interactions are 
included in the velocity $v_R$.
The second and third terms correspond to the intra- and
inter-edge Coulomb interactions respectively \cite{Moon96},
\begin{eqnarray}
S_a &=& {1\over 2}\int dxdy V_a(x-y)
\left\{
\rho_u(x)\rho_u(y)
+
\rho_l(x)\rho_l(y)
\right\}
\nonumber \\
S_w &=& \int dxdy V_w(x-y)\rho_u(x)\rho_l(y),
\nonumber
\end{eqnarray}
where
\[
V_a(x)={e^2\over\epsilon\sqrt{x^2+a^2}}\ ,\ 
V_w(x)={e^2\over\epsilon\sqrt{x^2+w^2}}
\]
with $a$ being an ultraviolet cutoff on the scale of lattice constant
and $\epsilon$ being the dielectric constant.
Bosonization of the densities on upper and lower edges are given by
\[
\rho_{u,l}(x)=
{1\over 2\pi}
{\partial\over\partial x}\phi_{u,l}(x)
\]
with $\phi_\pm=\phi_u\pm\phi_l$.
The last term in Eq.(1) describes a QPT through the bulk FQH liquid
$\psi_u^\dagger \psi_l+\psi_l^\dagger\psi_u$,
since the creation and annihilation of quasiparticles 
on the upper and lower edges are described respectively by the operators
$\psi_{u,l}=e^{\pm i\phi_{u,l}}$.
Integrating out $\phi_-$, we obtain an effective action for $\phi_+$, 
\begin{eqnarray}
S_0 +S_a+S_w
\nonumber \\
={1\over\beta}
\sum_{\omega}\int{dk\over 2\pi}
{v_R\over8\pi\nu}
\left\{
\eta_+ k^2
+{\omega^2\over\eta_- v_R^2}
\right\}
|\phi_+|^2,
\end{eqnarray}
where
$
\eta_\pm (k)=1+{\nu\over 2\pi}\{V_a(k)\pm V_w(k)\}
$
with 
$V_a(k)$ and $V_w(k)$ 
being the fourier transformation of 
$V_a(x)$ and $V_w(x)$.
Noting that $a \ll w$, we can evaluate $\eta_\pm(k)$ as
\begin{eqnarray}
\eta_+(k) &\sim& 1+\xi\ln{1\over|k|a},
\nonumber \\
\eta_-(k) &=& \left\{
\begin{array}{lr}
1+\xi\ln{w\over a}   & (|k|w \ll 1) \\
1+\xi\ln{1\over|k|a} & (|k|w \gg 1)
\end{array}
\right.
\nonumber
\end{eqnarray}
where
$\xi=(\nu\alpha/\pi\epsilon)(c/v)$
measures the strength of the inter-edge Coulomb interaction,
with $\alpha=e^2/\hbar c$ being the fine structure constant.
Eq. (2) gives the dispersion relation
$\omega(k) = v_R |k| \sqrt{ \eta_+ (k) \eta_-(k) }$ \cite{Morinari96,Wen95}.
The problem of the tunneling through a single barrier in TL liquids
was first studied by Kane and Fisher\cite{Kane92} and later extended
by Furusaki and Nagaosa\cite{Furusaki93}.
They derived an effective action for the phase field at the barrier site
by integrating out the continuum degrees of freedom.
\[
S_{QPT}[\theta]
= \frac{1}{4\pi\beta\nu}\sum_{\omega}
\zeta_w(\omega)|\omega| |\theta(\omega)|^2
+u\int d\tau\cos\theta(\tau),
\]
where $\theta$ is the phase at the point contact, and 
\[
\zeta_w(\omega) 
\sim \sqrt{
\frac{\eta_+(k_0)}{\eta_-(k_0)}}
= \left\{
\begin{array}{ll}
\sqrt{1+2\xi\ln{v_R\over\omega w}}
& (\omega<\Lambda_w)\\
1 & (\omega>\Lambda_w)
\end{array}
\right.
\]
with $\Lambda_w=v_R/w$ and $k_0=|\omega|/\sqrt{\eta_+(k_0)\eta_-(k_0)}$.

First we discuss the RG analysis at high temperatures,
we consider the limit where the scattering
potential is very weak.
We study the scaling behavior of the scattering potential
using the standard perturbative RG treatment~\cite{Kane92,Furusaki93}.
The scaling equation for $u$ is 
derived perturbatively by successively integrating over the
high frequency components, and the result is
\cite{Kane92,Furusaki93,Fisher85},
\begin{equation}
{d(u/\Lambda)\over u/\Lambda}
=\left(
{\nu\over\zeta_w(\Lambda)}-1
\right)
{d\Lambda\over\Lambda}.
\end{equation}
It turns out that as the cut-off $\Lambda$, 
which can be relpaced by the temperature $T$, the
scattering potential $u$ scales to stronger values, i.e., relevant.
The above perturbative treatment with respect to $u$
breaks down as $u/\Lambda$ becomes the order of unity,
where the crossover from weak to strong coupling occurs.
This crossover temperature $\Lambda_1$ is obtained by
setting $u(\Lambda_1)/\Lambda_1 = 1$ and is given by 
$\Lambda_1=\left({u_0/\Lambda_0^\nu}\right)^{1/(1-\nu)}$,
where
$u_0$ and $\Lambda_0$ are the bare strength of the potential
and cutoff, respectively.
Then the consideration here is restricted to the 
higher temperatures $T>\Lambda_1$.
When $\Lambda_1<\Lambda_w$,
$u(\Lambda)$ exhibits two different behaviors
corresponding to the two temperature regions.
Above $\Lambda_w$,
LRCI has no effect on the RG
equation, and $u(\Lambda) \propto \Lambda^{\nu-1}$.
The temperature dependence of $G(T)$ is 
obtained by the second order perturbation 
in the renormalized coupling constant $u(\Lambda)$
with the cut-off $\Lambda$ being replaced by the temperature $T$.
Then we obtain the usual temperature dependence
$G(T)-\nu{e^2/h}\sim - T^{2\nu-2}$.
When $\Lambda_1<\Lambda_w$,
there is the temperature region 
$\Lambda_1<T<\Lambda_w$,
where the solution of Eq.(3) is given by
\[
u(\Lambda)
=u(\Lambda_w)
\exp\left[
-{\nu\over \xi}
\left\{
\sqrt{1+2\xi\ln{\Lambda_w\over\Lambda}}-1
\right\}
\right],
\]
leading to the tunneling conductance 
\begin{eqnarray}
G(T) - \nu{e^2\over h}
\nonumber \\
\sim -{1\over T^2}
\exp\left[
-{ {2\nu} \over \xi}
\left\{
\sqrt{1+2\xi\ln{\Lambda_w\over T}}-1
\right\}
\right].
\end{eqnarray}
If we expand the exponent in Eq.~(4), the leading correction to
$G(T)-\nu{e^2\over h}$ is
\[
-\left({T\over\Lambda_w}\right)^{2\nu -2}
\exp\left[2\nu\xi
\left(\ln{\Lambda_w\over T}\right)^2
\right],
\]
which means that the LRCI further reduces $G(T)$.

Next we discuss the RG analysis at low temperatures,
where we consider the opposite limit where the scattering potential
is very strong.
This corresponds to the low temperature $T<\Lambda_1$.
In this limit, the electron transport can be viewed as the tunneling
of the phase $\theta$
from a potential minimum to an adjacent minimum.
This process corresponds to an instanton or an anti-instanton.
By the duality mapping in the dilute instanton gas approximation (DIGA),
we transform the original model
to an analogous model in the weak potential limit
~\cite{Schmid83},

\[
S_{DIGA}[\tilde{\theta}]
= \frac{\nu}{4\pi\beta}\sum_{\omega}
{|\omega|\over\zeta_w(\omega)}
|\tilde{\theta}(\omega)|^2
+2z\int d\tau\cos\tilde{\theta}(\tau),
\]
where $z$ is  the instanton fugacitiy, which is
the tunneling matrix elements from
$\theta=0$ to $\pm2\pi$.
It turns out that only when $\xi=0$,
which corresponds to the case of short range interactions,
the dual action can be identified with the original
one in terms of the correspondences
$\nu \leftrightarrow 1/\nu$,
$\tilde{\theta} \leftrightarrow \theta$
and
$2z \leftrightarrow u$.
According to the standard perturbative RG
treatment for the instanton
fugacity $z$, we obtain
\[
{d(z/\Lambda)
\over z/\Lambda}
=\left(
{\zeta_w(\Lambda)\over\nu}-1
\right)
{d\Lambda\over\Lambda}
. 
\]
$z(\Lambda)$ exhibits a crossover with a characteristic temperature $\Lambda_w$
when $\Lambda_w<\Lambda_1$.
In the region $\Lambda_w<T<\Lambda_1$,
the LRCI has no effect on the RG
equation.
When $T<\Lambda_w$,
the above RG equation can be solved in the same way as the
high temperature case.
We find that the tunneling of an instanton is
supressed by the LRCI. 
The conductance $G(T)$ is obtained in  the second order perturbation 
in the renormalized fugacity $z(\Lambda)$, 
i.e., the tunneling amplitude,
with the cut-off $\Lambda$ being replaced by $T$:
\[
G \sim
{1\over T^2}
\exp\left[-{2\over 3\nu\xi}
\left\{
\left(1+2\xi\ln{\Lambda_w\over T}\right)^{3/2}-1
\right\}\right],
\]
which has the asymptotic forms
for $ \Lambda_2 \ll T < \Lambda_w$,
\[
G \sim
T^{2/\nu-2}
\exp\left[-{\xi\over \nu}
\left(\ln{\Lambda_w\over T}\right)^2\right],
\]
while 
for $T \ll \Lambda_2$,
\[
G \sim  {1\over T^2}\exp\left[
-{4\over 3\nu}\sqrt{2\xi}\left(\ln{\Lambda_w\over T}\right)^{3/2}
\right],
\]
where $\Lambda_2=\Lambda_w e^{-1/2\xi}$.
At very low temperatures,
tunneling conductance decreases faster than any power law.
Here we evaluate the crossover temperatures.
We assume that bulk FQH liquid exhibits the $\nu=1/3$ plateau.
According to MG, we use $v_R\sim 10^5 m/s$,
$\xi\sim 0.2$, and $w\sim 60\mu m$.
In the QPT model these values give
$\Lambda_w\sim 10 mK$, and $\Lambda_2$ becomes the order of $1 mK$.
$\Lambda_1$ is controlled by the strength of the potential barrier
at the point contact.
Then we believe that the effect of LRCI in the QPT model
might resolve one of the discrepancies between the theoretical
prediction and the experiment pointed out by MG.\cite{Moon96}

\noindent
{\it Electron tunneling model}- Here we consider the model
where the bulk FQH liquid is divided into left and right condensates
with a characteristic separation $d \ll w$,
and electrons can tunnel between the left and right edges
through a insulating region at the point contact (Fig.~2).

We start with the case where $w$ is sufficiently large
compared to the energy scales in question.  
In this case left and right edges are assumed to be
pararell and infinitely long.
Experimentally such a situation can be realized by putting
a thin film insulator (with width $d \ll w$) between the FQH liquids.
Using this experimental geometry,
the rise of the tunneling conductance
will be observed experimentally.
We start with the effective action for the phase at the point contact $\theta$:
\begin{eqnarray}
S_{ET}[\theta]
&=& {1\over 4\pi\beta\nu}\sum_{\omega}
\zeta_d(\omega)|\omega| |\theta(\omega)|^2
\nonumber \\
&+&\gamma\int d\tau \cos {\theta(\tau)\over\nu},
\end{eqnarray}
where 
$\zeta_d(\Lambda)$ has the same form as $\zeta_w(\Lambda)$ with
$\Lambda_w$ replaced by $\Lambda_d=v_R/d$.
Here, note that the phase $\theta$ in the ET model has a mathematically
equivalent but physically different origin from the one in the QPT model,
due to the different oringins of two chiral TL liquids. 
The first term of Eq. (5) describes the edge modes of the
left and right FQH liquid with intra- and inter-edge Coulomb 
interactions.
The second term comes from ET between the left and right
edges:
$\Psi_L^{\dagger}\Psi_R+\Psi_R^{\dagger}\Psi_L
\sim\cos{\phi_+\over\nu}$,
where the electron operators on the
left and right edges are given by
$\Psi_{L,R}\sim\exp[ 
\pm i\phi_{L,R}/\nu]$
with $\phi_\pm=\phi_L\pm\phi_R$,
and $\gamma$ is the strength of ET.
In Eq.~(5), we have already integrated out the continuum
degrees of freedom.
Following the standard perturbative RG treatment, we obtain
\[
{d(\gamma/\Lambda)
\over
\gamma/\Lambda}
=\left(
{1\over\nu\zeta_d(\Lambda)}-1
\right)
{d\Lambda\over\Lambda}.
\]
As is the previous cases,
above the crossover temperature $\Lambda_d$,
LRCI has no effect on the RG
equation, i.e.
$G(T)\sim T^{2/\nu-2}$.
Below the crossover temperature $\Lambda_d$,
it is easy to see that
LRCI makes $G(T)$ decrease more
gradually than $T^{2/\nu-2}$.
What is more drastic is, however, that the ET becomes relevant
when the temperature is further lowered below
$\Lambda_3=\Lambda_d\exp[-({1/\nu^2}-1)/2\xi]$.
As the temperature is lowered from $\Lambda_d$,
$G(T)$ decreases more gradually than $T^{2/\nu-2}$,
and at $\Lambda_3$ it turns to the increase if $\Lambda_3>\Lambda_w$.

When the temperature is further
lowered below $\Lambda_w$, one has to take care of the 
edges extended to right and left rather than the 
edges facing to each other (Fig.~2).
To decibe the present situation
we start with the usual QPT model with LRCI.
After incorporating the LRCI, we throw away
the $x>0$ part and  
require the constraint that $\Psi_L+\Psi_R=0$ at $x=0$
for the QPT model, which means
$\phi_L(x,t)=-\phi_R(-x,t)+\nu\pi$
\cite{Eggert92}.
Thus we obtain the left branch of the ET model at
$T<\Lambda_w$.
Making the right branch in the same way,
we study the tunneling between them
to obtain the same low temperature dependence of $G(T)$
as the QPT model.

Now we can explicitly write down our prediction for the
tunneling conductance in the ET model.
As the temperature is lowered from $\Lambda_d  \gg \Lambda_w$,
$G(T)$ decreases more gradually than $T^{2/\nu-2}$,
and at $\Lambda_3$ it turns to the increase.
At lower temperatures than $\Lambda_d$, $G(T)$ scales as
\[
G(T)\sim
{\Lambda_d\over T^2}
\exp\left[-{2\over \nu\xi}
\left(\sqrt{1+2\xi\ln{\Lambda_d\over T}}-1\right)
\right]
\]
which have the following asymptotic form for $T \gg \Lambda_3$,
\[
G(T)\sim
T^{2/\nu-2}
\exp\left[{\xi\over \nu}
\left(\ln{\Lambda_d\over T}\right)^2\right],
\]
which means that the conductance is enhanced compared with the 
$T^{2/\nu-2}$ for the short-range interaction case.
In the region $\Lambda_w \ll T \ll \Lambda_3$, $G(T)$ scales
as $T^{-2}$.
At very low temperatures ($T<\Lambda_w$),
LRCI further reduces $G(T)$ as in the QPT model. 

Our treatment of ET model is
not equivalent to the one in MG \cite{Moon96},
where we believe that the charge phase and the Josephson phase 
are confused, although the final results are similar to ours. 
The Josephson pahse is a phase of 
$\Psi^{L\dagger}\Psi^{R\dagger}$,
and its gradient is proportional to the current and 
not to the density.
ET should be described as a cosine potential in terms of the
charge phase.

In summary, we study the effects of LRCI
on the quantum transport in FQH edges with $\nu=1/(2k+1)$.        
We consider two models, i.e., quasi-particle tunneling
(QPT) model and electron tunneling (ET) model.
Various crossovers of the tunneling conductance $G(T)$
as a function of the temperature $T$ are found.
In the QPT model the LRCI reduces the conductance $G(T)$
compared with the case of short range interaction.
In the ET model, on the other hand, there is a temperature
region where $G(T)$ is enhanced,
and even the nonmonotonic temperature dependence is possible.

The authors are grateful to A. Furusaki and T. Morinari
for useful discussions.
The work is supported by the Center of the Excellence.

\end{document}